\documentclass[prb,onecolumn,showpacs]{revtex4}
\usepackage{amssymb}
\usepackage{graphicx}

\input{tcilatex}

\begin{document}

\title{Charge dynamics in the phase string model for high-$T_{c}$
superconductors}
\author{Zheng-Cheng Gu and Zheng-Yu Weng}
\affiliation{\textit{Center for Advanced Study, Tsinghua University, Beijing 100084}}
\date{{\small \today}}

\begin{abstract}
An understanding of the anomalous charge dynamics in the high-$T_{c}$
cuprates is obtained based on a model study of doped Mott insulators. The
high-temperature optical conductivity is found to generally have a
two-component structure: a Drude like part followed by a mid-infrared band.
The scattering rate associated with the Drude part exhibits a
linear-temperature dependence over a wide range of high temperature, while
the Drude term gets progressively suppressed below a characteristic energy
of \emph{magnetic} origin as the system enters the pseudogap phase. The
high-energy optical conductivity shows a resonancelike feature in an
underdoped case and continuously evolves into a $1/\omega $ tail at higher
doping, indicating that they share the same physical origin. In particular,
such a high-energy component is closely correlated with the $\omega $-peak
structure of the density-density correlation function at \emph{different}
momenta, in systematic consistency with exact diagonalization results based
on the $t$-$J$ model. The underlying physics is attributed to the \emph{%
high-energy} spin-charge separation in the model, in which the
\textquotedblleft mode coupling\textquotedblright\ responsible for the
anomalous charge properties is \emph{not} between the electrons and some
collective mode but rather between new charge carriers, holons, and a novel
topological gauge field controlled by spin dynamics, as the consequence of
the strong short-range electron-electron Coulomb repulsion in the doped Mott
insulator.
\end{abstract}

\maketitle

\section{\protect\bigskip introduction}

Charge dynamics in the high-$T_{c}$ cuprates has been under intensive
studies for nearly two decades by now. With the continuous improvements in
experimental techniques and sample quality, the anomalous charge dynamics
exhibited in optical measurements has been firmly established as intrinsic
properties unique to the high-$T_{c}$ cuprates.

Some important features observed\cite{timusk,basov2} in the in-plane optical
measurements are: (a) The undoped parent compound is a Mott insulator with a
clear charge transfer gap observed in the optical conductivity. Upon doping,
spectral weight starts to appear inside the charge transfer gap.\cite{mid1}
(b) In heavily underdoped materials, the optical conductivity shows a very
clear two-component feature: a low-lying Drude component followed by a
resonancelike peak around the so-called high mid-infrared energy $\mathbf{%
\omega }$\textbf{$_{\mathrm{mir}}^{H}\sim $}$0.5-0.8$\textbf{\ }eV.\cite%
{mid1,mid2,basov1,low2} (c) Below some pseudogap temperature in the
underdoped cuprates, the low-energy part of the optical conductivity is
suppressed with decreasing temperature, accompanied by an emerging \emph{%
lower} mid-infrared resonancelike peak at $\omega _{\mathrm{mir}}^{L}$ ($%
\sim 0.1$ eV near the optimal case).\cite{low2,low1} (d) Near the optimal
doping, the high mid-infrared peak around $\omega _{\mathrm{mir}}^{H}$
continuously evolves into an approximate $1/\omega $ behavior in the optical
conductivity, with a shift of spectral weight towards lower energy.\cite%
{basov1,marginal1} At the same time, the scattering rate shows a linear-$%
\omega $ dependence,\cite{marginal2,qcp} while the scattering rate at $%
\omega \sim 0$ increases linearly with temperature, consistent with the dc
resistivity measurement.\cite{ong,resistivity1,ando}

Such anomalous electromagnetic response in the cuprates has posed a great
challenge to any microscopic theory with regard to the charge scattering
mechanism. Mainly focusing on the optimal case, the one-component
approaches, including the marginal Fermi-liquid theory,\cite{marginalFL}
projected Fermi liquid with edge singularities,\cite{LL} slave-boson gauge
theory,\cite{pali1} nearly antiferromagnetic Fermi-liquid theory,\cite%
{AFFL,SF} etc., have all stressed how the scattering rate $1/\tau $ may
become strongly frequency-dependent as the consequence of strong
interactions. Conventional electron-boson-mode coupling mechanisms have been
also suggested in this regard.\cite{mode3,mode4,mode5,mode6} Alternatively
the power-law behavior of the scattering rate $1/\tau (\omega )\sim \omega
^{\alpha }$ has also been interpreted as due to some zero-temperature
quantum critical point hiding below the superconducting dome in the phase
diagram.\cite{qcp1}

To fit the underdoped data, however, a multi-component model composed of a
free-carrier Drude term and a set of Lorentzian oscillators seems more
appropriate based on a phenomenological consideration to account for the
high mid-infrared band.\cite{timusk,basov2} Note that the presence of a
mid-infrared band has been indeed found in various microscopic models
describing the hole motion in a quantum antiferromagnet.\cite%
{holeAF1,holeAF2,holeAF3} On the other hand, numerical studies\cite{dagotto2}
based on the $t$-$J$ model and Hubbard model have also shown the existence
of a mid-infrared absorption at small doping and its evolution into an
approximately $1/\omega $ tail with increasing doping. These all point to
the fact that strong correlations may be essential in order to fully
understand the overall charge dynamics in the cuprates.

In this paper, we will study the global feature of charge dynamics based on
an effective theory describing a doped Mott insulator. Due to the strong
correlation nature, the charge scattering mechanism is no longer a
conventional mode coupling between the electrons and some collective bosonic
mode. It is replaced by a new type of \textquotedblleft mode
coupling\textquotedblright , in which the \emph{spinless} charge carriers,
holons, interact with a topological gauge field, with the latter controlled
by the \emph{neutral} spin dynamics. Namely the strong short-range Coulomb
interaction effect between the electrons is now represented by a novel
scattering between the charge and spin degrees of freedom of the system.

Base on this model, we show that the high-temperature optical conductivity
is generically associated with a two-component behavior. Here the low-$%
\omega $ Drude-like term is characterized by a linear-temperature-dependent
scattering rate over a wide range of temperature, which gets progressively
suppressed as the system enters into the pseudogap phase at low temperature
with the emerging of a lower mid-infrared peak. On the other hand, the high-$%
\omega $ mid-infrared part exhibits a resonancelike feature in the strong
scattering case and continuously evolves into a $1/\omega $ tail in the weak
scattering limit, which can be related to the underdoping and high doping,
respectively. The high-energy mid-infrared component is further shown to be
closely correlated with the $\omega $-peak structure of the density-density
correlation function at \emph{different} momenta which is systematically
consistent with the exact diagonalization results based on the $t$-$J$
model. Therefore, we establish a consistent picture for the charge dynamics
in the cuprate superconductors in different temperature and energy regimes
within a \emph{single} unified theoretical framework.

The rest of paper will be organized as follows. In Sec. II, the basic model
describing a doped Mott insulator will be introduced, and various important
temperature and energy scales decided by the model will be discussed. In
particular, a novel charge scattering mechanism embedded in the model will
be emphasized. In Sec. III, we study the high-temperature charge dynamics
where the gauge fluctuations are reduced to static to allow for a numerical
simulation. The optical conductivity and scattering rate are calculated. The
two-component $\omega $ structure and its correlation with the
density-density correlation function will be discussed in detail. In Sec.
IV, the scattering strength is substantially reduced in the pseudogap phase
where a perturbation approach is allowed to study the low-energy optical
conductivity. A pseudogap behavior of the optical conductivity with the
emergence of a lower mid-infrared resonance will be determined here.
Finally, a summary and discussions will be presented in Sec. V.

\section{basic model}

From a doped-Mott-insulator point of view, the $t$-$J$ model is believed to
be the simplest relevant model for the high-$T_{c}$ cuprates. Finding the
correct low-energy effective theory for the $t$-$J$ model has been a
fascinating focus in search for the mechanism of high-$T_{c}$
superconductivity. The concepts of the resonating-valence-bond (RVB) \cite%
{anderson1} and spin-charge separation \cite{anderson2} have been
conjectured in order to understand the strong correlation nature of the
electrons in such a system, although, different from the one-dimensional
case,\cite{exact1,1dexp} no well-defined spinless charge carriers (holons)
and $S=1/2$ neutral spin excitations (spinons) have ever been unambiguously
identified experimentally in the two-dimensional case. Nevertheless, there
has been strong numerical evidence indicating that the charge and spin
dynamics should be described separately by two different degrees of freedom
with distinct characteristic energy scales.\cite{exact1,exact2}

Microscopic studies of the $t$-$J$ model indicate that residual strong
correlations still generally exist, in a spin-charge separation description,
between the two building blocks -- spinless \textquotedblleft
holons\textquotedblright\ and charge neutral \textquotedblleft
spinons\textquotedblright . For example, the interaction is mediated by the
U(1) \cite{U(1)} or SU(2) \cite{SU(2)} gauge field in the slave-boson
approach. In contrast, in the phase string theory\cite{ps} the elementary
force between the bosonic \textquotedblleft holons\textquotedblright\ and
\textquotedblleft spinons\textquotedblright\ is mediated by the mutual
Chern-Simons gauge fields. These gauge interactions complicate the issue
regarding how to make a clear identification of holons and spinons, even if
they exist in the cuprates.

Nevertheless, the presence of a spin-charge separation will predict a set of
highly nontrivial charge and spin dynamics whose unique features can be
subjected to a systematic comparison with the experimental measurement. Even
in low-temperature phases where the confinement of \textquotedblleft
holons\textquotedblright\ and \textquotedblleft spinons\textquotedblright\
due to the gauge fields may occur \emph{at low energies}, the composite
structure of the electrons in terms of \textquotedblleft
holons\textquotedblright\ and \textquotedblleft spinons\textquotedblright\
should still exhibit interesting features with the increase of energy, which
approaches the \textquotedblleft asymptotic freedom\textquotedblright\ at
some shorter distance and higher energy. An experimental probe of
high-energy excitations is particularly meaningful here.

In the following, we explore the charge dynamics based on an effective gauge
theory derived from the $t$-$J$ model and show how the spin-charge
separation exhibits and plays the essential role there.

\subsection{Phase string model}

Our starting point is the phase string model,\cite{ps,weng98,wavefunction}
which takes the form: $H_{\mathrm{string}}=H_{h}+H_{s}$ with 
\begin{eqnarray}
H_{h} &=&-t_{h}\sum_{\left\langle ij\right\rangle }\left(
e^{iA_{ij}^{s}+eA_{ij}^{e}}\right) h_{i}^{\dagger }h_{j}+H.c.+\lambda
_{h}\sum_{i}\left( n_{i}^{h}-\delta \right)  \label{md1} \\
H_{s} &=&-J_{s}\sum_{\left\langle ij\right\rangle \sigma }\left( e^{i\sigma
A_{ij}^{h}}\right) b_{i\sigma }^{\dagger }b_{j-\sigma }^{\dagger
}+H.c.+\lambda \sum_{i\sigma }\left[ n_{i\sigma }^{b}-(1-\delta )/2\right]
\label{md3}
\end{eqnarray}%
in which the holon field, described by the bosonic annihilation operator $%
h_{i},$ carries the full charge $+e$ coupling to the external
electromagnetic vector potential $A_{ij}^{e}$.

The spinless holon field and the neutral $S=1/2$ spinon field (which carries
the spin index as described by the bosonic annihilation operator $b_{i\sigma
})$ are minimally coupled to two internal gauge field $A_{ij}^{s}$ and $%
A_{ij}^{h}$, whose gauge-invariant strengths are constrained to the matter
fields by the following relations, respectively 
\begin{eqnarray}
\sum_{C}A_{ij}^{s} &=&\pi \sum_{l\in \Sigma _{C}}(n_{l\uparrow
}^{b}-n_{l\downarrow }^{b})  \label{cond1} \\
\sum_{C}A_{ij}^{h} &=&\pi \sum_{l\in \Sigma _{C}}n_{l}^{h}  \label{cond2}
\end{eqnarray}%
where $n_{l\sigma }^{b}=b_{l\sigma }^{\dagger }b_{l\sigma }$ and $%
n_{l}^{h}=h_{l}^{\dagger }h_{l}$, respectively, denote the on-site spinon
and holon number operators, and $\Sigma _{C}$ is the region enclosed by an
arbitrary (counterclockwise) closed loop $C.\,$Such relations are known as
the 
mutual Chern-Simons gauge structure which dictate that a holon and a spinon
perceive each other as a fictitious $\pi $-flux tube [here to avoid the
short-distance uncertainty at each center of a $\pi $-flux tube, on the
right-hand sides (rhs) of Eqs. (\ref{cond1}) and (\ref{cond2}), the
distribution of a holon or spinon at site $l$ should be understood as being
slightly smeared within a small area centered at $l$]$.$

The phase string model thus defined explicitly respects all the symmetries
including the translational, time-reversal, parity, and spin rotational
symmetries. Note that the spin operators in this model are defined by $%
S_{i}^{z}=1/2\sum_{\sigma }\sigma n_{i\sigma }^{b}$, $S_{i}^{+}=$ $%
b_{i\uparrow }^{\dagger }b_{i\downarrow }(-1)^{i}e^{i\Phi _{i}^{h}}$, and $%
S_{i}^{-}=\left( S_{i}^{+}\right) ^{\dag },$ where $\Phi _{i}^{h}-\Phi
_{j}^{h}=2A_{ij}^{h}$ with the core of each flux-tube being smeared within a
small area as mentioned above.

In Eqs. (\ref{cond1}) and (\ref{cond2}), $\lambda _{h}$ and $\lambda $ are
the chemical potentials to implement $\sum_{l}n_{l}^{h}=N\delta $ and $%
\sum_{l\sigma }n_{l\sigma }^{b}=N(1-\delta )\equiv N\bar{n}^{b}$, where $N$
is the total number of lattice sites and $\delta $ the doping concentration.
The effective hopping integral $t_{h}\sim 0.67t$\cite{wavefunction} and in
this paper we shall choose $t_{h}=2J$ which corresponds to $t\sim 3J$. The
renormalized superexchange coupling $J_{s}=J\left\langle \hat{\Delta}%
^{s}\right\rangle (1-2g\delta )/2$ ($g\sim 2$)\cite{psnmr} with $J_{s}\sim
J/2$ at low doping$,$ where $t$ and $J$ are the bare parameters in the
original $t$-$J$ model. Here the gauge-invariant bosonic RVB order parameter 
$\Delta ^{s}=\left\langle \hat{\Delta}^{s}\right\rangle \equiv \left\langle
\sum_{\sigma }e^{-i\sigma A_{ij}^{h}}b_{i\sigma }b_{j-\sigma }\right\rangle $
is determined self-consistently.

\subsection{Phase diagram and basic energy scales}

\begin{figure}[tbp]
\begin{center}
\includegraphics[width=3.5in]
{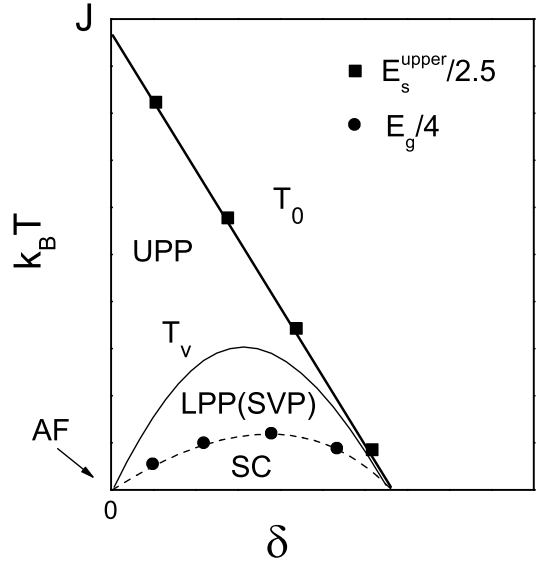}
\end{center}
\caption{In the phase diagram of the phase string model, the characteristic
temperature $T_{0}$ for the upper pseudogap phase (UPP) and $T_{c}$ for the
superconducting phase (SC) are shown to be well scaled with two basic energy
scales, $E_{s}^{\mathrm{upper}}$ and $E_{g}$ (see Fig. \protect\ref{chi} for
their definitions), with the proportional coefficients $\sim 0.4$ and $0.25$%
, respectively. The lower pseudogap phase (LPP) or the spontaneous vortex
phase (SVP) is characterized by a temperature $T_{v}$ which is also
determined by the spin excitation spectrum.\protect\cite{pslpp1}}
\label{phasediagram}
\end{figure}

Fig. \ref{phasediagram} shows the basic phase diagram\cite{psnmr,pslpp1} for
the phase string model given in Eqs. (\ref{md1}) and (\ref{md3}). The
characteristic temperature $T_{0}$ denotes the boundary of the so-called the
upper pseudogap phase (UPP), described by the bosonic RVB order parameter $%
\Delta ^{s}\neq 0$ based on $H_{s}$;\cite{psnmr} The so-called lower
pseudogap phase (LPP) or spontaneous vortex phase (SVP) is nested within the
UPP with a characteristic temperature $T_{v}$, which corresponds to the
bosonic degenerate regime for the holons;\cite{pslpp1,pslpp2} Finally the
superconducting (SC) phase coherence is realized inside the LPP/SVP at lower
temperatures. In Fig. \ref{phasediagram} both LPP and SC phase terminate at
half-filling where an antiferromagnetic (AF) long-range order is recovered
in the ground state of $H_{s}$. Note that a more careful study with
considering the longer-range AF correlations will lead to the vanishing of
the SC transition temperature $T_{c}$ at a finite critical doping
concentration $x_{c}.$\cite{pscritical} 

\begin{figure}[tbp]
\begin{center}
\includegraphics[width=3.5in]
{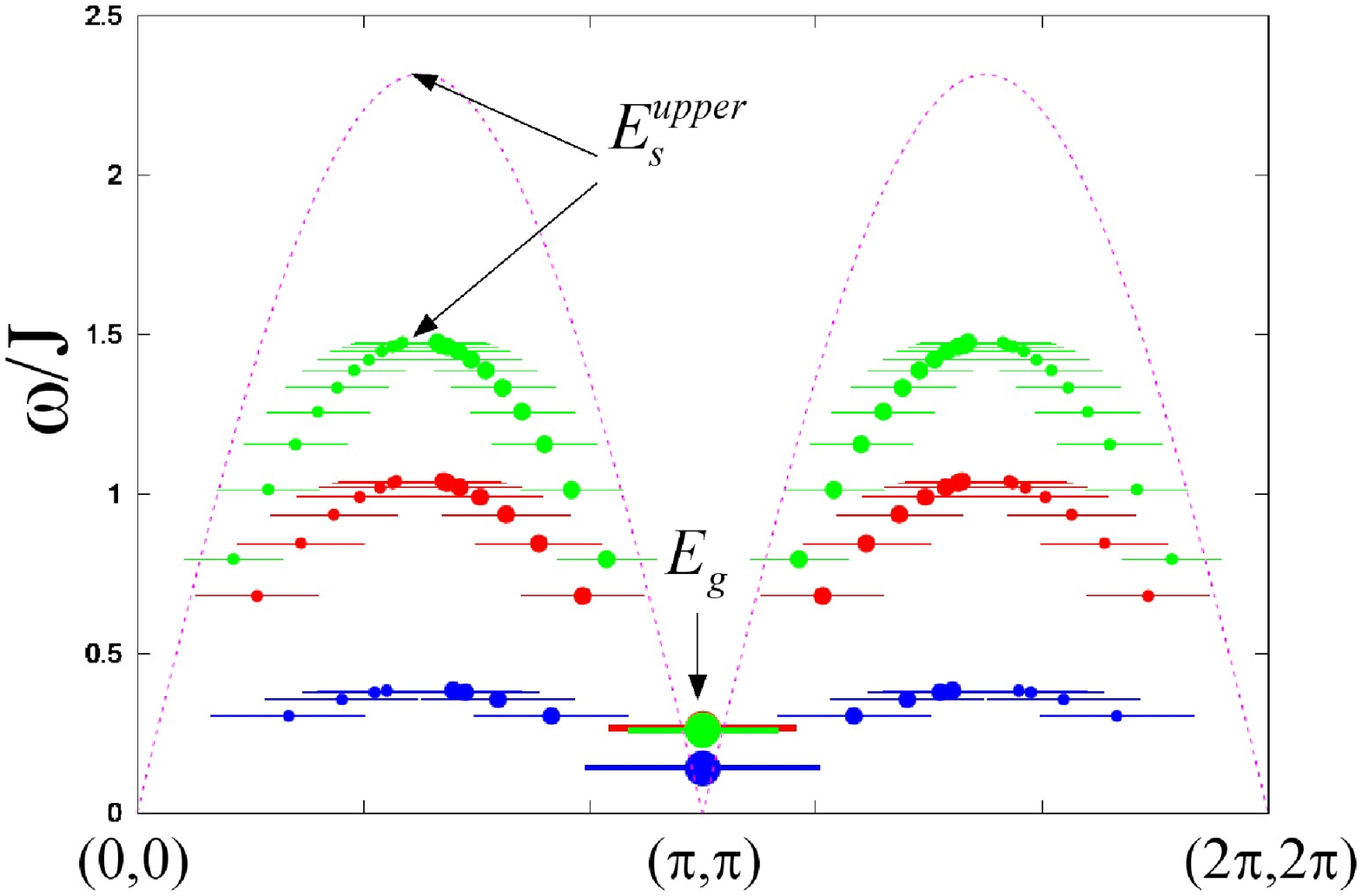}
\end{center}
\caption{Basic energy scales, $E_{s}^{\mathrm{upper}}$ and $E_{g}$,
determined by the dynamic spin susceptibility function $\mathrm{Im}\protect%
\chi ^{zz}(\mathbf{q},\protect\omega )$ at $T=0$. The peak positions of $%
\mathrm{{Im}\protect\chi ^{zz}}$ at $\protect\delta =0$ is shown in the
energy and momentum (along the $q_{x}=q_{y}$ axis) by the dotted curve,
which tracks the spin wave dispersion with $E_{s}^{\mathrm{upper}}\simeq
2.3J $ and $E_{g}=0$. The upper-bound energy $E_{s}^{\mathrm{upper}}$
monotonically decreases with increasing doping from $\ 0.05$, $0.125$, to $%
0.2$. $E_{g}$ denotes the resonancelike peak energy at $\mathbf{q}=(\protect%
\pi ,\protect\pi )$, which emerges in the LPP and SC state at finite doping.
Note that the finite horizontal bars at finite doping indicate the momentum
widths for these non-propagating modes.\protect\cite{psneutron}}
\label{chi}
\end{figure}

Physically the UPP is a regime where short-range AF correlations start to
develop quickly with decreasing temperature$.$\cite{psnmr} At half-filling,
it continuously evolves into the AF long-range ordered state at $T=0$, where
the gapless spin-wave excitation is shown in Fig. \ref{chi} by the dotted
curve.\cite{psneutron} However, such long-range AF correlations get
\textquotedblleft truncated\textquotedblright\ at finite doping after
entering LPP, where an energy gap $E_{g}$ is opened up around the AF wave
vector $\mathbf{Q}=(\pi ,\pi )$ as shown in Fig. \ref{chi}, which depicts
the peak positions of the spin dynamic susceptibility function $\mathrm{Im}%
\chi ^{zz}(\mathbf{q},\omega )$ at $T=0$\cite{psneutron} (see Appendix A).

The doping dependence of the resonancelike spin energy $E_{g}$ is
illustrated in Fig. \ref{phasediagram} which scales with $T_{c}$ as $%
T_{c}\simeq E_{g}/4k_{\mathrm{B}}$ as previously established\cite{shaw}
based on the phase string model. The peak structure of $\mathrm{Im}\chi
^{zz}(\mathbf{q},\omega )$ in Fig. \ref{chi} coincides with the spin-wave
spectrum at half-filing (dotted curve) and becomes a non-propagating mode
with a finite, doping-dependent width in the momentum space at finite
doping. But the overall high-energy envelop of the latter still well tracks
that of the (softened) spin wave, implying that the short-range AF
correlations remain strong in the short-distance, high-energy regime in the
UPP. This is consistent with the neutron scattering measurements.\cite%
{neutron} There is a characteristic \emph{upper-bound} energy scale, $E_{s}^{%
\mathrm{upper}}$, indicated by the arrows in Fig. \ref{chi} for the $S=1$
excitations. Its doping dependence follows closely with $T_{0}$ in Fig. \ref%
{phasediagram} and monotonically increases with reducing doping, reaching $%
E_{s}^{\mathrm{upper}}\simeq 2.3J$ at half-filling.

The low-lying sharp resonancelike structure in $\mathrm{Im}\chi ^{zz}(%
\mathbf{q},\omega )$ at $E_{g}$ is the consequence\cite{psneutron} of the 
\emph{holon condensation} realized in $H_{h}$ which then influences Eq. (\ref%
{md3}) via the gauge field $A_{ij}^{h}$ in terms of Eq. (\ref{cond2}). With
the increase of temperature, the holon condensation (or more precisely the
amplitude condensation of the holon field) will terminate, not at $T_{c}$,
but at an intermediate characteristic temperature scale $T_{v}$ lying
between $T_{0}$ and $T_{c}$ that represents the onset of the LPP/SVP in Fig. %
\ref{phasediagram}. Such a bosonic degenerate regime corresponds to the
Nernst regime observed in the cuprates.\cite{nernst}

\subsection{Novel scattering mechanism}

The charge dynamics is governed by $H_{h}$ [Eq. (\ref{md1})], in which a
bosonic holon, carrying the full charge $+e$, will get scattered off by an
internal gauge field $A_{ij}^{s}$. According to Eq. (\ref{cond1}), an
isolated spinon excitation will serve as a $\pi $-flux tube perceived by the
holons and thus provides a strong, unconventional charge scattering source.
So at high temperature when a lot of spinons are thermally excited, one
expects a severe intrinsic frustration effect exerted from $A_{ij}^{s}$ on
the holons. On the other hand, the effect of the $\pi $-flux tubes bound to
those spinons which are RVB paired at short-distance will be essentially
cancelled out. At sufficiently low temperature and low energy where the
majority of spins remain short-range RVB paired, the gauge fluctuations in $%
A_{ij}^{s}$ will then be substantially reduced to result in a weak
scattering, which warrants a perturbation treatment.

To see how the spin dynamics influences the charge degree of freedom via $%
A_{ij}^{s}$, one may introduce $\mathbf{A}^{s}\cdot \left( \mathbf{r}_{i}%
\mathbf{-r}_{j}\right) \equiv A_{ij}^{s}$ and express the propagator of the
gauge field $\mathbf{A}^{s}$ in the continuum limit as follows 
\begin{eqnarray}
D_{\alpha \beta }^{A^{s}}(\mathbf{q},i\omega _{n}) &\equiv &\int_{0}^{\beta
}d\tau e^{i\omega _{n}\tau }\langle T_{\tau }A_{\alpha }^{s}(\mathbf{q,}\tau
)A_{\beta }^{s}(-\mathbf{q,0})\rangle  \nonumber \\
&=&-\left( \delta _{\alpha \beta }-\frac{q_{\alpha }q_{\beta }}{q^{2}}%
\right) \frac{4\pi ^{2}}{q^{2}a^{4}}\chi ^{zz}(\mathbf{q},i\omega _{n})
\label{pp}
\end{eqnarray}%
using Eq. (\ref{cond1}), where $\chi ^{zz}(\mathbf{q},i\omega _{n})$ is the $%
\hat{z}$-component spin susceptibility function$.$

The detailed spin dynamics described by $H_{s}$ in Eq. (\ref{md3}) has been
systematically studied\cite{psneutron,psnmr} before. For example, the peak
structure in $\mathrm{Im}\chi ^{zz}(\mathbf{q},\omega )$ at the mean-field
level\cite{psneutron} ($T=0$) is presented in Fig. \ref{chi} at various
dopings. At finite doping, due to the opening up of the low-lying spin gap $%
E_{g}$, one finds a vanishing spectral weight of \textrm{Im}$D^{A^{s}}(%
\mathbf{q},\omega )$ at $\omega <E_{g}$ and $T=0$ according to Eq. (\ref{pp}%
). Since the spin gap $E_{g}$ is directly related to the longest-size RVB
pairs of spins, the effect of $A^{s},$ contributed by the $\pi $ flux tubes
bound to individual spinons in terms of Eq. (\ref{cond1}), will get
cancelled out cleanly in such a low-energy, long-wavelength regime.
Consequently the bosonic holons in Eq. (\ref{md1}) will be free from the
flux frustration and thus experience a Bose condensation at a sufficiently
low temperature, which defines $T_{v}$ for the\ LPP as previously stated. In
fact, the zero-temperature spin dynamic susceptibility in Fig. \ref{chi} is
obtained based on $H_{s}$ under the holon condensation. Namely, the spin gap
and holon condensation actually enforce each other self-consistently in the
phase string model.

The gauge fluctuations in $D^{A^{s}}(\mathbf{q},\omega )$ will gain a finite
strength at $E_{g}\leq \omega \leq E_{s}^{\mathrm{upper}}$ and then fall off
rapidly beyond $E_{s}^{\mathrm{upper}}$ based on $\mathrm{Im}\chi ^{zz}(%
\mathbf{q},\omega )$. At $\omega \gg E_{s}^{\mathrm{upper}}$, the holons
will mainly experience quasistatic gauge fluctuations concentrated at $%
\omega \lesssim $ $E_{s}^{\mathrm{upper}},$ with the total strength
determined by \ 
\begin{equation}
\left\langle \left( \Phi _{{\small \square }}^{s}\right) ^{2}\right\rangle
=\int d\omega \frac{1}{N}\sum_{\mathbf{q}}4\pi ^{2}S^{zz}(\mathbf{q},\omega )
\label{phi}
\end{equation}%
where $\Phi _{{\small \square }}^{s}=a^{2}\mathbf{\hat{z}\cdot }\left(
\nabla \times \mathbf{A}^{s}\right) $ defines the local flux per plaquette
(surrounding a lattice site) and the spin structure factor $S^{zz}(\mathbf{q}%
,\omega )=\pi ^{-1}\left[ 1+n\left( \omega \right) \right] \mathrm{Im}\chi
^{zz}(\mathbf{q},\omega )$. According to the sum rule discussed in Ref.\cite%
{psneutron}, one finds 
\begin{equation}
\sqrt{\left\langle \left( \Phi _{_{\square }}^{s}\right) ^{2}\right\rangle }%
\simeq \pi \sqrt{\frac{(3-\delta )(1-\delta )}{3}}\equiv \left\vert \Phi
_{_{\square }}^{s}\right\vert _{\mathrm{\max }}  \label{pi}
\end{equation}

In particular, if the temperature is further increased to $T\gtrsim T_{0}$,
there is no more significant AF correlations among spins as $\Delta ^{s}=0$
such that (see Appendix A) 
\begin{equation}
S^{zz}(\mathbf{q},\omega )=\frac{1}{4\pi ^{2}}\left\vert \Phi _{_{\square
}}^{s}\right\vert _{\mathrm{\max }}^{2}\delta (\omega )  \label{static}
\end{equation}%
Then the corresponding gauge flux fluctuation becomes truly \emph{static }%
with the weight $\sqrt{\left\langle \left( \Phi _{_{\square }}^{s}\right)
^{2}\right\rangle }$ concentrating at $\omega =0$.

Namely, the gauge field $A^{s}$ simply describes the randomly distributed
static flux at $T\geq T_{0}$ and quasistatic flux at high energy $\omega \gg
E_{s}^{\mathrm{upper}}$ in the case of $T<T_{0}$. Such regimes represent the
maximal quantum frustration that the holons can experience in the phase
string model governed by Eq. (\ref{md1}).

\section{Charge dynamics at high temperature: Two-component feature}

In Sec. II C, we have seen that the strongest scattering to the charge
carriers in the phase string model will set in at $T\geq T_{0}$ above the
UPP, where the fluctuations of the gauge field $\mathbf{A}^{s}$ are
concentrated at $\omega =0$, i.e., in the static limit. In the following we
shall study the charge response as characterized by the optical conductivity
and density-density correlation function in this regime. It has been noted
that even at $T<T_{0}$, if $\omega \gg $ $E_{\mathrm{upper}}^{s}$, a
quasi-static approximation for the gauge field $A_{ij}^{s}$ should also be
justified. In other words, the results obtained at $T\geq T_{0}$ should be
still qualitatively applicable at $T<T_{0}$ at sufficiently \emph{high
energies}.

\subsection{The density of states for holons}

In the phase string model, the charge holons are coupled to the spin degrees
of freedom via the topological gauge field $A_{ij}^{s}$ defined in Eq. (\ref%
{cond1}). Thus the charge dynamics is strongly influenced by the spin
dynamics which decides the propagator of $A_{ij}^{s}$ in terms of Eq. (\ref%
{pp}). As discussed in Sec. II C, the spin fluctuations are greatly reduced
in the UPP at $T\geq T_{0}$ and become \textquotedblleft
static\textquotedblright\ [Eq. (\ref{static})] at the mean-field level as $%
\Delta ^{s}=0$. Accordingly, $A_{ij}^{s}$ describes a static random flux at $%
T\geq T_{0}$ with the strength given by Eqs. (\ref{phi}) and (\ref{static}).

Namely, above the UPP, the holons in $H_{h}$ simply see a collection of $\pm
\pi $ flux-tubes bound to spinons of number $(1-\delta )N$ which are
de-paired ($\Delta ^{s}=0)$ and randomly distributed in this regime with
neglecting the spin-spin correlation at the mean-field level.

Then, at $T\geq T_{0}$ one can make a direct numerical simulation to
diagonalize $H_{h}$ for each random configuration of spinons like an
impurity problem by a unitary transformation $h_{i}=\sum_{m}C_{im}a_{m}$
with $C_{im}$ satisfying%
\begin{equation}
-t_{h}\sum_{j=NN(i)}e^{iA_{ij}^{s}}C_{jm}=\varepsilon _{m}^{h}C_{im}
\label{C}
\end{equation}%
where $j=NN(i)$ denotes the four nearest neighbor sites. Note that the
hard-core interaction between the holons is neglected here as we are mainly
interested in the high-temperature or high-energy charge behavior below.

For the simplicity of numerical simulation, we define the smeared flux
strength of $A_{ij}^{s}$ on each \emph{lattice} plaquette by 
\begin{equation}
\sum_{\mathrm{plaquette}}A_{ij}^{s}=\Phi _{_{\square }}  \label{fluxpl}
\end{equation}%
and assume $\Phi _{_{\square }}=\pm \left\vert \Phi _{_{\square
}}\right\vert _{\max }$ with the signs $+$ and $-$ randomly distributed on
the lattice plaquette and 
\begin{equation}
\left\vert \Phi _{_{\square }}\right\vert _{\max }=(1-\delta )\pi
\label{fluxm}
\end{equation}%
Note that this assumption is valid when the individual $\pi $-flux tubes of
total number of $1-\delta $ spinons are well distinguished at high energy,
short-distance scales, and $\left\vert \Phi _{_{\square }}\right\vert _{\max
}$ is compatible in magnitude with $\left\vert \Phi _{_{\square
}}^{s}\right\vert _{\max }$ in Eq. (\ref{pi}) of the mean-field version. But
it should be emphasized that $\left\vert \Phi _{_{\square }}\right\vert
_{\max }$ in Eq. (\ref{fluxm}) only provides an upper bound for the average
flux strength per plaquette. One may think of many effects that can reduce
its magnitude. For example, two flux tubes of opposite signs sitting at the
nearest sites can partially cancel each other; the residual weak AF\ spin
correlations among spinons will further enhance such cancellation; in
addition, at larger doping the hopping of holons is always accompanied by a
\textquotedblleft backflow\textquotedblright\ of spinons which can mix $\pm
\pi $ fluxes and reduce the average strength of $\left\vert \Phi _{_{\square
}}\right\vert $ from $\left\vert \Phi _{_{\square }}\right\vert _{\max }$.

\begin{figure}[tbp]
\begin{center}
\includegraphics[width=3.5in]
{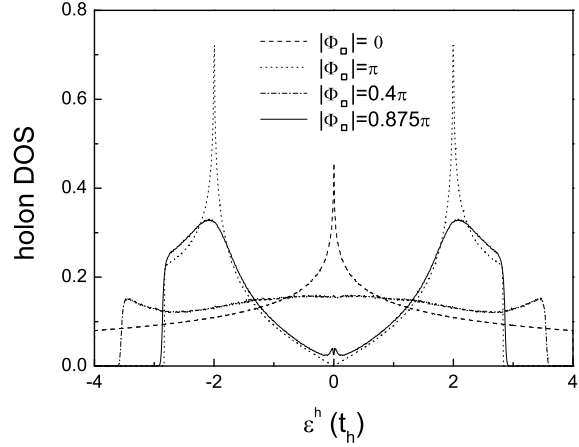}
\end{center}
\caption{Solid curve: the holon density of states (DOS) determined by $%
H_{h}\,$with the flux strength decided by Eq. (\protect\ref{fluxm}) at $%
\protect\delta =0.125$ (see text) and the dash-dotted curve: with the
reduced flux strength $\left\vert \Phi _{_{\square }}\right\vert =0.4\protect%
\pi $. By comparison, the dashed curve represents the flux-free limit, while
the dotted curve corresponds to the case in the presence of uniform $\protect%
\pi $ flux per plaquette. The calculation is done on a $32\times 32$
lattice. }
\label{DOS}
\end{figure}

Fig. \ref{DOS} illustrates the holon density of states (DOS) obtained at $%
\delta =0.125$ (solid curve). Here we have used the \emph{quenched} method
to average static random flux configurations of $\Phi _{_{\square }}.$ The
result shows how the DOS drastically reshaped by the gauge field, i.e., the 
\emph{suppression} in the high-energy (mid-band) DOS, as compared to the
flux free case (dashed curve). Note that the dotted curve in Fig. \ref{DOS}
represents the DOS for the case of a uniform $\pi $ flux per plaquette,
which looks similar to the present random flux case except that the momenta
remains well defined in a reduced Brillouin zone in contrast to a strong
mixing of momenta over a wide range by the scattering effect in the latter.
By comparison, the DOS with a reduced $\left\vert \Phi _{_{\square
}}\right\vert _{\max }=0.4\pi $ (dash-dotted curve) is also presented. It is
noted that other kinds of treatment for the static gauge field $A_{ij}^{s}$
will generically lead to the similar overall behavior. For example, one may
treat $\Phi _{_{\square }}$ as a white-noise random flux distributed within
the interval $\left[ -\left\vert \Phi _{_{\square }}\right\vert _{\max
},+\left\vert \Phi _{_{\square }}\right\vert _{\max }\right] $ or even
introduce some spatial correlation based on Eq. (\ref{pp}). 

\subsection{Optical conductivity: Two-component structure}

Let us consider the optical conductivity using the Kubo formula 
\begin{equation}
\sigma _{xx}(\mathbf{q=0},\omega )=\frac{i}{\omega }\left[ \Pi _{xx}(\mathbf{%
q=0},\omega )-e^{2}\langle T_{x}\rangle \right]  \label{sigma}
\end{equation}%
where $\Pi _{xx}(\mathbf{q=0},\omega )\equiv $ $\Pi _{xx}(i\omega
_{n}\rightarrow \omega +i0^{+})$ is the retarded current-current correlation
function with $\Pi _{xx}(i\omega _{n})=-\int_{0}^{\beta }d\tau e^{i\omega
_{n}\tau }\langle T_{\tau }J_{x}(\tau )J_{x}(0)\rangle $ and the charge
current operator defined by 
\begin{equation}
J_{x}(\mathbf{q})=\frac{iet_{h}}{\sqrt{N}}\sum_{i}e^{-i\mathbf{q}\cdot 
\mathbf{r}_{i}}\left( h_{i+\hat{x}}^{\dagger }h_{i}e^{iA_{i+\hat{x}%
,i}^{s}}-h_{i}^{\dagger }h_{i+\hat{x}}e^{iA_{i,i+\hat{x}}^{s}}\right)
\label{Jx}
\end{equation}%
$\langle T_{x}\rangle $ in Eq. (\ref{sigma}) is defined by 
\begin{equation}
\langle T_{x}\rangle =\frac{1}{N}\left\langle -t_{h}\sum_{i}e^{-i\mathbf{q}%
\cdot \mathbf{r}_{i}}\left( h_{i+\hat{x}}^{\dagger }h_{i}e^{iA_{i+\hat{x}%
,i}^{s}}+h.c.\right) \right\rangle  \label{Tx}
\end{equation}

The real part of the optical conductivity is then given by 
\begin{equation}
\sigma _{xx}^{\prime }(\omega )=\frac{\pi }{N}\sum_{m,m^{\prime
}}M_{mm^{\prime }}\frac{n(\xi _{m})-n(\xi _{m^{\prime }})}{\xi _{m^{\prime
}}-\xi _{m}}\delta (\omega -\xi _{m^{\prime }}+\xi _{m})  \label{op6}
\end{equation}%
in which $n(\xi _{m})=1/(e^{\beta \xi _{m}}-1)$ is the Bose distribution
factor with $\xi _{m}\equiv \varepsilon_{m}^{h}+\lambda _{h}$ and the matrix
element%
\begin{equation}
M_{mm^{\prime }}\equiv e^{2}t_{h}^{2}\left\vert
\sum_{i}(e^{iA_{i+x,i}^{s}}C_{i+x,m}^{\ast }C_{i,m^{\prime
}}-e^{iA_{i,i+x}^{s}}C_{i,m}^{\ast }C_{i+x,m^{\prime }})\right\vert ^{2}
\label{M}
\end{equation}%
with using Eq. (\ref{C}) for a given static configuration of $A_{ij}^{s}$.
Note that the final $\sigma _{xx}^{\prime }(\omega )$ is obtained by
averaging over the quenched random flux configurations as discussed above.

\begin{figure}[tbp]
\begin{center}
\includegraphics[width=3.5in]
{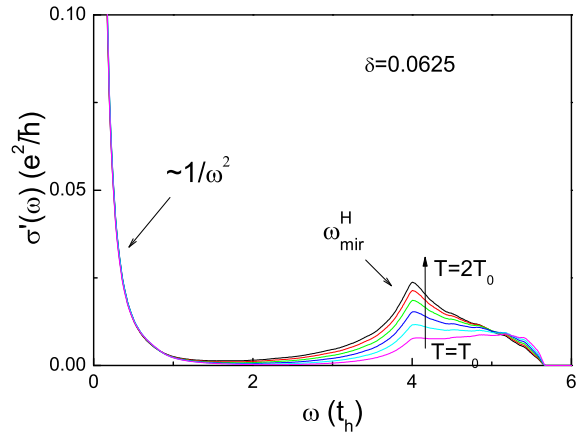}
\end{center}
\caption{The real part of the optical conductivity at $\protect\delta %
=0.0625 $ with $|\Phi _{{\protect\small \square }}|=|\Phi _{{\protect\small %
\square }}|_{\max }$ and $T_{0}\sim 0.5t_{h}.$}
\label{optical1}
\end{figure}

\begin{figure}[tbp]
\begin{center}
\includegraphics[width=3.5in]
{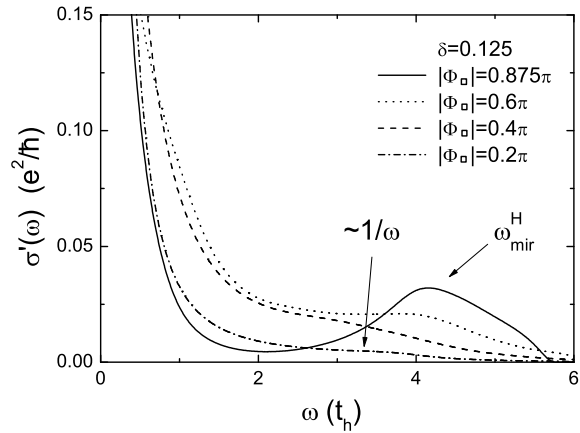}
\end{center}
\caption{The real part of the optical conductivity at $\protect\delta =0.125$
with $|\Phi _{{\protect\small \square }}|$ chosen from $|\Phi _{%
{\protect\small \square }}|_{\max }=0.875\protect\pi $ to $0.2\protect\pi $
at a fixed $T=0.5t_{h}$. }
\label{optical2}
\end{figure}

Fig. \ref{optical1} shows the calculated real part of the optical
conductivity at $\delta =0.065$, with choosing the maximal flux strength in
terms of Eq. (\ref{fluxm}). The main feature of the spectral curves at
various temperatures $T\geq T_{0}$ is that they can all be decomposed into a
two-component structure with a usual low-energy Drude component ($\sim
1/\omega ^{2}$) and a mid-infrared resonancelike peak around the energy
scale $\omega _{\mathrm{mir}}^{H}\sim 4t_{h}$, with the whole spectrum
eventually terminated below the energy $6t_{h}$.

\subsubsection{Mid-infrared peak}

The origin of the mid-infrared resonance has been one of the most intriguing
optical properties in the underdoped cuprates.\cite{timusk,basov1,basov2}
Normally a photon with the momentum $\mathbf{q}\sim 0$ cannot excite a
high-energy particle-hole pair involving a large momentum transfer due to
the momentum conservation law, unless there is a scattering mechanism to
strongly and widely smear the momentum. This is difficult to realize in a
conventional electron-collective-mode coupling mechanism. The present model
provides an alternative scattering mechanism due to the strong correlation
effect caused by the on-site Coulomb repulsion in a doped Mott insulator.

We have already seen that the effect of $A^{s}$ results in a double-peak
structure in the holon DOS (Fig. \ref{DOS}). In contrast to the uniform $\pi 
$ flux case shown in the same figure, which also has a double-peak
structure, the high-energy inter-peak transition at $\mathbf{q}\rightarrow 0$
becomes possible in the random flux case due to the mixing between the small
and large momenta by the strong scattering via $A^{s}.$ This is the origin
for the mid-infrared peak found in Fig. \ref{optical1}.

The presence of a peak in the optical conductivity around $\omega \sim 3t-4t$
has been previously identified in the Hubbard and $t$-$J$ models by exact
numerical simulations\cite{dagotto1,dagotto2} Two approaches are consistent
in this regard. Note that for such a high-energy and short-distance physics
the finite-size effect in the exact diagonalization calculation should not
be important. In Sec. III C below, we shall further compare the
density-density correlation function obtained by the exact diagonalization
and present approach, where the $\omega $-structures at different momenta
can provide much richer features in further support of the consistency.

The above-discussed mid-infrared peak in the optical conductivity seems in
contrast with the approximate $1/\omega $ behavior observed experimentally
for the optimally and over-doped cuprates.\cite{basov1} But if one
artificially reduces the magnitude $\left\vert \Phi _{_{\square
}}\right\vert $ from $\left\vert \Phi _{_{\square }}\right\vert _{\max }$
given in Eq. (\ref{fluxm}), the mid-infrared peak will actually smoothly
evolve into the $1/\omega $ behavior with reducing $\left\vert \Phi
_{_{\square }}\right\vert $, as clearly illustrated in Fig. \ref{optical2}
at a fixed holon concentration $\delta =0.125$. In Fig. \ref{optical2}, the
mid-infrared resonancelike peak at smaller $\left\vert \Phi _{_{\square
}}\right\vert $'s becomes softened and finally behaves like a $1/\omega $
tail in the regime $\sim 2t_{h}-4t_{h}$ with the weight shifting towards the
lower energy. As explained before, the strength of $\left\vert \Phi
_{_{\square }}\right\vert $ is expected to decrease faster with increasing
doping than $\left\vert \Phi _{_{\square }}\right\vert _{\max }$ given in
Eq. (\ref{fluxm}). Such a picture is also consistent with the exact
diagonalization calculations in the $t$-$J$ model.\cite{exact3}

\subsubsection{Low-energy component}

Now let us closely examine the low-$\omega $ component of the optical
conductivity. Experimentally the scattering rate is normally defined by 
\begin{equation}
\frac{1}{\tau (\omega )}=\left( \frac{\omega _{p}^{2}}{4\pi }\right) \mathrm{%
Re}\left[ \frac{1}{\sigma (\omega )}\right]  \label{1tao}
\end{equation}%
which is determined by the measured optical conductivity. Here $\omega _{p}$
denotes the plasma frequency, which in the present case is given by $\omega
_{p}=\sqrt{8\pi e^{2}\delta t_{h}}$. 
\begin{figure}[tbp]
\begin{center}
\includegraphics[width=3.5in]
{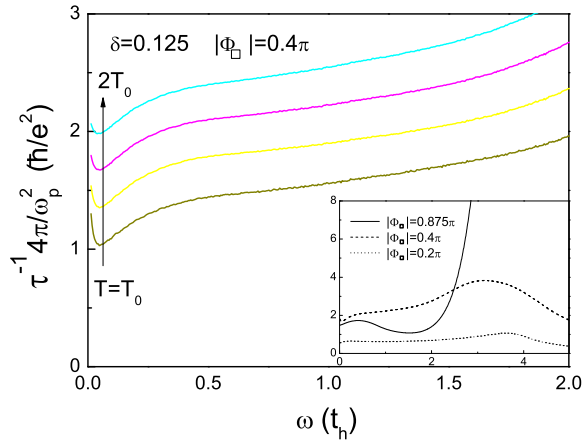}
\end{center}
\caption{The scattering rate $1/\protect\tau (\protect\omega )$ defined by
Eq. (\protect\ref{1tao}) at various temperatures between $T_{0}\simeq
0.25t_{h}$ and $2T_{0}$ which show a rough linear-$\protect\omega $
dependence over a wide range at $\protect\omega >k_{\mathrm{B}}T_{0}$.
Inset: $1/\protect\tau (\protect\omega )$ vs. $\protect\omega $ at different 
$\left\vert \Phi _{{\protect\small \square }}\right\vert $'s corresponding
to Fig. \protect\ref{optical2}. }
\label{tao}
\end{figure}
\begin{figure}[tbp]
\begin{center}
\includegraphics[width=3.5in]
{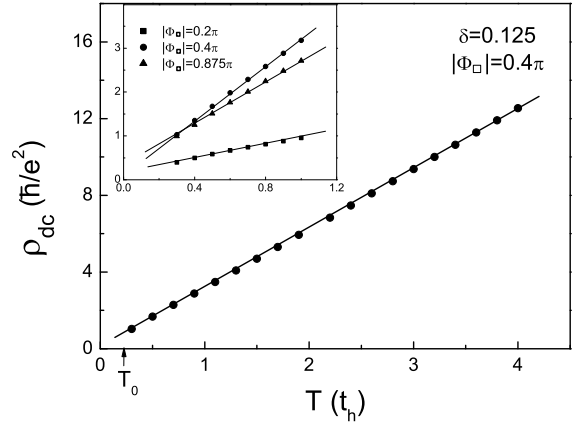}
\end{center}
\caption{The dc resistivity $\protect\rho _{\mathrm{dc}}=1/\protect\sigma %
^{\prime }(\protect\omega )|_{\protect\omega \sim 0}$ as a function of
temperature which is fit by a straight line showing the linear-$T$
dependence. Inset: $\protect\rho _{\mathrm{dc}}$ at different $\left\vert
\Phi _{{\protect\small \square }}\right\vert $'s which all show good linear-$%
T$ behavior with slightly different slopes.}
\label{resistivity}
\end{figure}

In Fig. \ref{tao}, $1/\tau (\omega )$ based on the calculated $\sigma
(\omega )$ is plotted as a function of $\omega $ in different temperatures
at $\delta =0.125$. Here we have chosen $\left\vert \Phi _{_{\square
}}\right\vert =0.4\pi $ which corresponds to the case where the high-$\omega 
$ optical conductivity looks more like a $1/\omega $ behavior (Fig. \ref%
{optical2}). Here one finds that $1/\tau (\omega )$ increases monotonically
with $\omega $ and is roughly linear-$\omega $ dependent over a wide $\omega 
$ region at $\omega >k_{\mathrm{B}}T_{0}$. Note that generally the $\omega $%
-dependence of $1/\tau (\omega )$ at higher energies is closely correlated
with the evolution of the aforementioned mid-infrared feature, as shown in
the inset of Fig. \ref{tao}. Especially, as shown in the inset of Fig. \ref%
{tao}, the non-monotonic behavior of $1/\tau (\omega )$ is also found in the
strong gauge fluctuation case (e.g., $\left\vert \Phi _{_{\square
}}\right\vert _{\max }=0.875\pi $), which corresponds to the heavily
underdoped case, and is consistent with recent experimental results.\cite%
{basov2,basov1}

In particular, one sees a parallel shift of $1/\tau (\omega )$ with
increasing temperature at low-$\omega $ in Fig. \ref{tao}, which implies a
linear-temperature dependence of the dc scattering rate. Such a parallel
shift is also observed in the experiments\cite{qcp,temp1} which persists to
a very high energy scale ($\sim 3000cm^{-1}$), implying an unconventional
scattering mechanism in such strongly correlated systems. The dc scattering
rate $1/\tau _{\mathrm{dc}}$ can be determined by extrapolating $1/\tau
(\omega )$ to $\omega =0$. Due to the parallel shift, the temperature
dependence of $1/\tau (\omega )$ at low-$\omega $ is quite similar. The
obtained dc resistivity based on the Drude formula $\rho _{\mathrm{dc}%
}=\left( \omega _{p}^{2}/4\pi \right) \tau _{\mathrm{dc}}^{-1}$ $=1/\sigma
^{\prime }(0)$ is shown in Fig. \ref{resistivity} which is indeed quite
linear over a very wide range of temperature at $T\geq T_{0}$.

In order to understand the physical origin of the linear-$T$ behavior, one
may rewrite $\sigma ^{\prime }(0)$ in terms of Eq. (\ref{op6}) as follows%
\[
\sigma ^{\prime }(0)=\beta \frac{\pi }{N}\sum_{m}n(\xi _{m})[1+n(\xi _{m})]%
\left[ \sum_{m^{\prime }}M_{mm^{\prime }}\delta (\xi _{m}-\xi _{m^{\prime }})%
\right] 
\]%
where the chemical potential $\lambda _{h}$ in $\xi _{m}$ is determined by $%
\sum_{m}n(\xi _{m})=N\delta $. We find $\sigma ^{\prime }(0)\propto $ $\beta 
$ over a very wide range of the temperature at $T>T_{0}$ where $n(\xi
_{m})\ll 1,$ i.e., in the classical regime of the bosons. The corresponding
scattering rate $\hslash /\tau _{\mathrm{dc}}\sim 0.7k_{B}T$ for the case
shown in the main panel of Fig. \ref{resistivity}, whose slope is slightly $%
\left\vert \Phi _{_{\square }}\right\vert $ dependent as indicated in the
inset. Indeed, as discussed in Ref. \cite{pslpp1}, the bosonic degenerate
regime already ends up at $T_{v}$, i.e., the boundary of the LPP/SVP (Fig. %
\ref{phasediagram}), where the excited spinon number becomes equal to the
holon number such that the quantum phase coherence among the latter get
totally interrupted by the former which carry $\pi $-flux tubes. At $T\geq
T_{0}$, totally $1-\delta $ randomly distributed $\pi $-flux tubes are
perceived by $\delta $ holons and the latter behave like classical
particles. One expects this anomalous transport be smoothly connected to the
Brinkman-Rice retracing path regime\cite{br} in the large $T$ limit.

The dc scattering rate $\hslash /\tau _{\mathrm{dc}}\sim 2k_{B}T$ has been
previously obtained\cite{pali3} by the quantum Monte Carlo numerical method,
where the starting model is a system of interacting bosons coupled with
strong Gaussian fluctuations of the static gauge field of the strength $%
\left\langle \left( \Phi _{{\small \square }}^{s}\right) ^{2}\right\rangle .$
Note that $\left\langle \left( \Phi _{{\small \square }}^{s}\right)
^{2}\right\rangle $ used in the Monte Carlo simulation is about the same
order of magnitude as in the above case and, in particular, it is
temperature independent in contrast to the linear-$T$ dependence predicted
in the slave-boson \textrm{U(1) }gauge theory\cite{pali1} which was the
original motivation for such a Monte Carlo study.\cite{pali3} It is noted
that the short-range repulsion between the holons has been taken into
account in the Mont Carlo calculation, which may be responsible for the
larger temperature slope of the scattering rate as compared to the present
approach besides the Gaussian approximation used there. A further difference
in the treatment is that we have used a quenched method to average over the
static random flux configurations of $\Phi _{_{\square }}$ since $A_{ij}^{s}$
depicts $\pi $-flux tubes bound to random distributed spinons at $T\geq
T_{0} $, while an \emph{annealing} approximation is used in Ref. \cite{pali3}
because of the Gaussian fluctuations of the gauge flux employed there. As we
are mainly interested in the high-temperature (or high-energy) behavior,
such a difference is qualitatively not important as discussed in Ref. \cite%
{pali2}.

It is noted that the lattice effect becomes very important at high-energy,
short-distance at such strong flux fluctuations. It is actually responsible
for the double-peak holon DOS, the mid-infrared feature in the optical
conductivity, and the high-$\omega $ behavior of $1/\tau (\omega )$
discussed above. In the following we examine an another consequence of this
unique DOS structure.

\subsection{Density-density correlation function}

In the above section, the mid-infrared resonance peak of the $\mathbf{q}=0$
optical conductivity has been attributed to a large-$\omega $ transition
between the double peaks of the holon DOS [Fig. \ref{DOS}], which is the
consequence of the holon coupling with the strong fluctuating gauge field.
In the following we discuss an independent probe of such a peculiar DOS
structure by studying the density-density function at finite momentum $%
\mathbf{q}$ and energy $\omega ,$ and compare the results with the exact
numerical calculations.

The charge (holon) number operator in the momentum space is expressed as 
\begin{equation}
n(\mathbf{q})=\sum_{i}e^{i\mathbf{q}\cdot \mathbf{r}_{i}}h_{i}^{\dagger
}h_{i}  \label{nq}
\end{equation}%
Similar to the (retarded) current-current correlation function, the
imaginary part of the (retarded) density-density correlation function can be
expressed as 
\[
C_{d}(\mathbf{q},\omega )=\frac{\pi }{N}\sum_{m,m^{\prime }}\left\vert
\sum_{i}e^{i\mathbf{q}\cdot \mathbf{r}_{i}}C_{i,m}^{\ast }C_{i,m^{\prime
}}\right\vert ^{2}\left[ n(\xi _{m})-n(\xi _{m^{\prime }})\right] \delta
(\omega -\xi _{m^{\prime }}+\xi _{m}) 
\]%
Corresponding to the calculated optical conductivity in Figs. \ref{optical1}
and \ref{optical2}, using the same parameters and method, the structure
function $C_{d}(\mathbf{q},\omega )$ of the density-density correlation can
be similarly computed. 
\begin{figure}[tbp]
\begin{center}
\includegraphics[width=3.5in]
{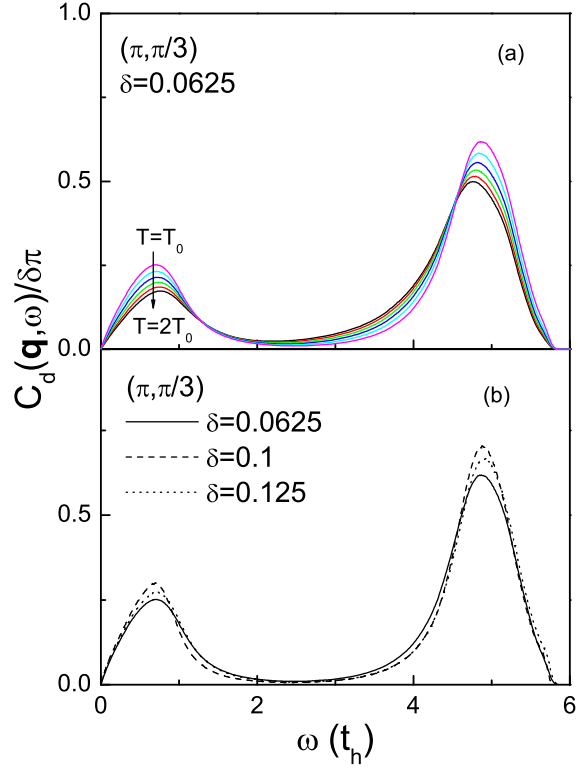}
\end{center}
\caption{(a) The density-density correlation function $C_{d}(\mathbf{q},%
\protect\omega )/\protect\delta \protect\pi $ at $\mathbf{q=}$ $(\protect\pi %
,\protect\pi /3)$ at $\protect\delta =0.0625$ at various temperatures; (b)
The scaling behavior of $C_{d}(\mathbf{q},\protect\omega )/\protect\delta 
\protect\pi $ at different dopings at $T=0.5t_{h}$. }
\label{scaling}
\end{figure}
\begin{figure}[tbp]
\begin{center}
\includegraphics[width=4in]
{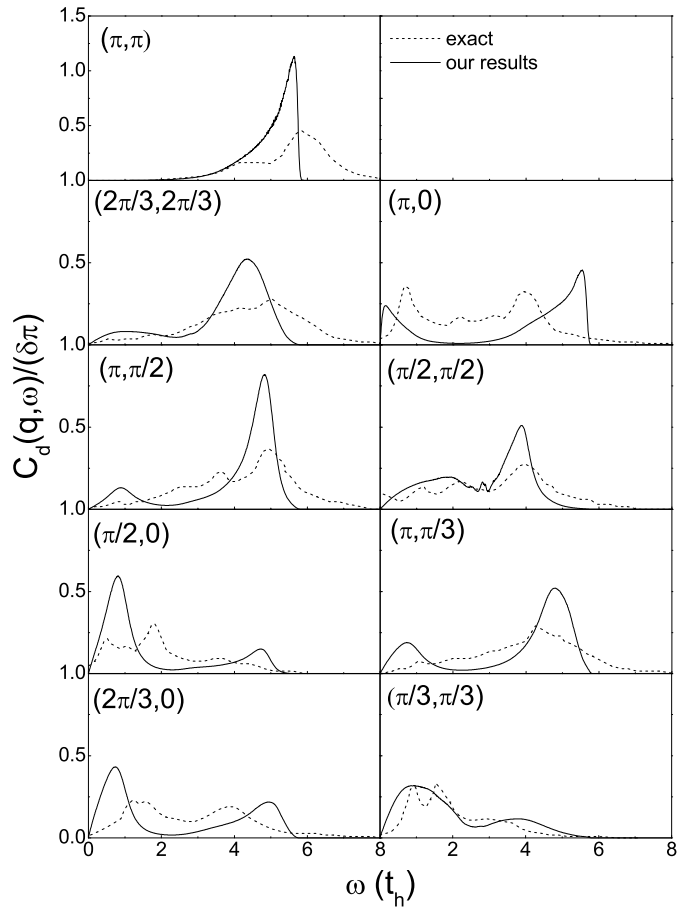}
\end{center}
\caption{The calculated density-density correlation function $C_{d}(\mathbf{q%
},\protect\omega )/\protect\delta \protect\pi $ at different momenta (solid
curves) with $T=0.5t_{h}$. The exact diagonalization results\protect\cite%
{exact2} are shown as dashed curves.}
\label{density}
\end{figure}

The calculated $C_{d}(\mathbf{q},\omega )/\delta \pi $ at momentum $\mathbf{q%
}=(\pi ,\pi /3)$ are presented in Fig. \ref{scaling} at different
temperatures and doping concentrations. The double-peak structure as a
function of $\omega $ reflects the underlying structure of the DOS of holons
in Fig. \ref{DOS} with $\Phi _{_{\square }}=\pm \left\vert \Phi _{_{\square
}}\right\vert _{\max }$. An interesting feature shown in Fig. \ref{scaling}%
(b) is that $C_{d}(\mathbf{q},\omega )/\delta \pi $ is roughly
doping-independent. Such a \textquotedblleft scaling\textquotedblright\
behavior was previously found numerically for the $t$-$J$ model,\cite{exact2}
which is generally inconsistent with the picture of particle-hole
excitations in a Fermi liquid and was conjectured to be an indication of the
bosonic description of charge excitations.\cite{exact2}

The calculated $C_{d}(\mathbf{q},\omega )/\delta \pi $ is presented in Fig. %
\ref{density} (solid curves), which evolves distinctively with different
momenta. For comparison, the exact diagonalization results\cite{exact2} are
presented as dotted curves. Due to the approximate doping-independence, the
numerical result of four holes in a $4\times 4$ lattice is used here. It is
interesting to see that the overall $\omega $-peak feature of the calculated
density-density correlation function is in qualitative and systematic
agreement with the numerical one at different $\mathbf{q}$'s without fitting
parameters (here $t$ is simply set at $t_{h}$ as the mid-infrared feature
peaks around $\sim 4t$ in the numerical calculation). Such a consistency
between the present effective theory and the exact diagonalization provides
an another strong evidence, in addition to the mid-infrared resonance peak
of the same origin, that the gauge-coupling boson model (\ref{md1})
correctly captures the high-energy charge excitations in the $t$-$J$ model
and large-$U$ Hubbard model.

\section{Low-Energy Pseudogap Behavior at Low Temperature}

So far we have been focused on the \textquotedblleft normal
state\textquotedblright\ above the UPP at $T\geq T_{0}$ where we have seen
the maximal \emph{static} scattering coming from the gauge field. As
outlined in Sec. II, when the temperature decreases below $T_{0}$, the spins
start to form bosonic RVB pairs and develop short-range AF correlations.
Although this does not change the overall integrated strength of the gauge
fluctuations, due to the spin sum rule as given in Eqs. (\ref{phi}) and (\ref%
{pi}), the low-$\omega $ gauge fluctuations of $\mathbf{A}^{s}$ will get
suppressed progressively. Namely the scattering felt by the low-lying charge
carriers will be reduced. Here the gauge fluctuations are no longer static
and are strongly correlated with the spin dynamics, where the above static
approximation method is not applicable at least in the low-energy regime$.$

In the following we shall study the low-energy optical conductivity at $%
T<T_{0}$ by using the perturbative method and continuum approximation
instead. Such an approach is meaningful in the regime where the spin
fluctuations are substantially suppressed, which in turn results in the weak
fluctuations of $\mathbf{A}^{s}$ according to Eq.\ (\ref{pp}). Of course, as
emphasized before, in the high-energy regime ($\omega \gg E_{s}^{\mathrm{%
upper}})$ the charge carriers still feel the strong scattering by a
quasistatic gauge fluctuation of the overall strength given by Eq. (\ref{pi}%
) and the previously results at $T\geq T_{0}$ are expected to still hold
qualitatively.

\subsection{Self-energy}

In the case of weak gauge fluctuations, one can take the continuum limit in
the Hamiltonian (\ref{md1}) as follows 
\begin{equation}
H_{h}=-\frac{1}{2m_{h}}\int d^{2}\mathbf{r}h^{\dagger }(\mathbf{r})\left[ 
\mathbf{\triangledown }-i\mathbf{A}^{s}(\mathbf{r})\right] ^{2}h(\mathbf{r}%
)+\lambda _{h}\int d^{2}\mathbf{r}h^{\dagger }(\mathbf{r})h(\mathbf{r})
\end{equation}%
where $m_{h}=\left( 2t_{h}a^{2}\right) ^{-1}$ with a shift in the chemical
potential $\lambda _{h}-4t_{h}\rightarrow \lambda _{h}$. Further express $%
H_{h}=H_{0}+H_{1}$ with%
\begin{eqnarray}
H_{0} &=&\int \frac{d^{2}\mathbf{k}}{{(2\pi )}^{2}}\left( \frac{k^{2}}{2m_{h}%
}+\lambda _{h}\right) h_{\mathbf{k}}^{\dagger }h_{\mathbf{k}}  \label{self2}
\\
H_{1} &=&-\frac{1}{2m_{h}}\left[ \int \frac{d^{2}\mathbf{k}}{{(2\pi )}^{2}}%
\int \frac{d^{2}\mathbf{k}^{\prime }}{{(2\pi )}^{2}}(\mathbf{k}+\mathbf{k}%
^{\prime })\cdot \mathbf{A}^{s}(\mathbf{q})+\int \frac{d^{2}\mathbf{q}_{1}}{{%
(2\pi )}^{2}}\mathbf{A}^{s}(\mathbf{q}_{1})\cdot \mathbf{A}^{s}(\mathbf{q}%
_{2})\right] h_{\mathbf{k}}^{\dagger }h_{\mathbf{k}^{\prime }}
\end{eqnarray}%
in which $\mathbf{q}=\mathbf{k}-\mathbf{k}^{\prime }$ and $\mathbf{q}_{1}+%
\mathbf{q}_{2}=\mathbf{k}-\mathbf{k}^{\prime }$.

\begin{figure}[tbp]
\begin{center}
\includegraphics[width=3in]
{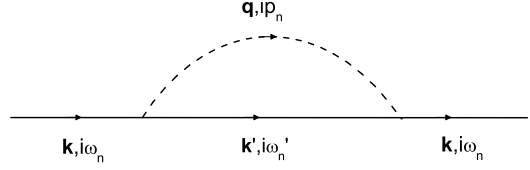}
\end{center}
\caption{Feynman diagram for the boson self-energy $\Sigma (\mathbf{k},i%
\protect\omega _{n}).$ }
\label{digram1}
\end{figure}

We treat $H_{1}$ as a perturbation and calculate the holon self-energy up to
the quadratic order of the gauge field. In the imaginary-time
representation, the self-energy can be written as 
\begin{equation}
\Sigma (\mathbf{k},i\omega _{n})=-\frac{1}{\beta (2m_{h})^{2}}%
\sum_{ip_{n}}\int \frac{d^{2}\mathbf{k}^{\prime }}{{(2\pi )}^{2}}G^{(0)}(%
\mathbf{k}^{\prime },i\omega _{n}^{\prime })D_{\alpha \beta }^{A^{s}}(%
\mathbf{q},ip_{n})(\mathbf{k}+\mathbf{k}^{\prime })^{\alpha }(\mathbf{k}+%
\mathbf{k}^{\prime })^{\beta }
\end{equation}%
where the free boson field propagator $G^{(0)}(\mathbf{k},i\omega
_{n})=1/\left( i\omega _{n}-\xi _{\mathbf{k}}\right) $ with $\xi _{\mathbf{k}%
}=\frac{\mathbf{k}^{2}}{2m_{h}}+\lambda _{h}$. The corresponding Feynman
diagram is illustrated in Fig. \ref{digram1} with $\mathbf{q}=\mathbf{k}-%
\mathbf{k}^{\prime }$ and $p_{n}=\omega _{n}-\omega _{n}^{\prime }$. 
\begin{figure}[tbp]
\begin{center}
\includegraphics[width=3.5in]
{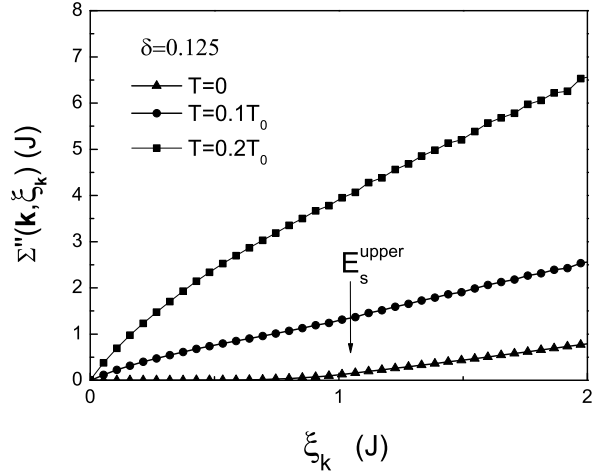}
\end{center}
\caption{The imaginary holon self-energy at various temperatures below $%
T_{0} $.}
\label{selfenergy}
\end{figure}
By using the spectral function representation of $D^{A^{s}}$ in terms of $%
\mathrm{Im}\chi ^{zz}$ according to Eq. (\ref{pp}) and after the frequency
summation and analytic continuation $i\omega _{n}\rightarrow \omega +i0^{+}$%
, one finally obtains the imaginary part of the self-energy 
\begin{equation}
\Sigma ^{\prime \prime }(\mathbf{k},\omega )=-\frac{4\pi ^{2}}{(2m_{h})^{2}}%
\int_{-\infty }^{\infty }d\omega ^{\prime }\int \frac{d^{2}\mathbf{k}%
^{\prime }}{(2\pi )^{2}}\mathrm{Im}\chi ^{zz}(\mathbf{q},\omega ^{\prime })%
\frac{|\mathbf{k}\times \mathbf{q}|^{2}}{q^{4}}\left[ n(\xi _{\mathbf{k}%
^{\prime }})+n(\omega ^{\prime })+1\right] \delta (\omega -\xi _{\mathbf{k}%
^{\prime }}-\omega ^{\prime })
\end{equation}

Based on the dynamic spin susceptibility function $\mathrm{Im}\chi ^{zz}$
determined\cite{psneutron} by $H_{s}$ (see Appendix A), $\Sigma ^{\prime
\prime }(\mathbf{k},\xi _{\mathbf{k}})$ can be then numerically computed and
the results are presented in Fig. \ref{selfenergy} at several
low-temperatures below $T_{0}$. It shows that generally $\Sigma ^{\prime
\prime }(\mathbf{k},\xi _{\mathbf{k}})\propto \xi _{\mathbf{k}}$ at high
energy with the slope dependent on temperature. But at low energy, $\Sigma
^{\prime \prime }(\mathbf{k},\xi _{\mathbf{k}})$ is quickly suppressed below 
$E_{s}^{\mathrm{upper}}$ at $T=0$ or reduced from the linear-$\epsilon _{%
\mathbf{k}}^{h}$ behavior at finite temperature due to the presence of spin
dynamics, characterized by $\mathrm{Im}\chi ^{zz}$ shown in Fig. \ref{chi},
which makes the total strength of the gauge fluctuations spread over a
finite $\omega $ region such that the effective scattering becomes weakened
at low energy.

However, in contrast to a fermion system, the self-energy $\Sigma ^{\prime
\prime }(\mathbf{k},\xi _{\mathbf{k}})$ is not generally related to the
measurable transport properties for the present boson system. So in the
follow we shall directly consider the optical conductivity based on the Kubo
formula.

\subsection{Optical conductivity}

The optical conductivity $\sigma ^{\prime }(\omega )$ is determined by the
retarded current-current correlation function. Note that in the continuum
limit, the gauge-invariant current in the phase string model is given by 
\begin{equation}
\mathbf{J}=-\frac{ie}{m_{h}}\int d^{2}r[h^{\dagger }(\mathbf{r})\nabla h(%
\mathbf{r})-i\mathbf{A}^{s}(\mathbf{r})h(\mathbf{r})^{\dagger }h(\mathbf{r})]
\end{equation}%
which can be further written in two parts $\mathbf{J}=\mathbf{J}^{A}+\mathbf{%
J}^{B}$, with%
\begin{eqnarray}
\mathbf{J}^{A} &=&\frac{e}{m_{h}}\sum_{\mathbf{k}}\mathbf{k}h_{\mathbf{k}%
}^{\dagger }h_{\mathbf{k}} \\
\mathbf{J}^{B} &=&-\frac{e}{m_{h}\sqrt{N}a}\sum_{\mathbf{k},\mathbf{q}}%
\mathbf{A}^{s}(\mathbf{q})h_{\mathbf{k+q}}^{\dagger }h_{\mathbf{k}}
\label{opc3b}
\end{eqnarray}%
Correspondingly, to leading order of approximation, the current-current
correlation function contains two parts: $\Pi \equiv \Pi ^{A}+\Pi ^{B}$,
where

\begin{eqnarray}
\Pi ^{A}(i\omega _{n}) &=&-\frac{e^{2}}{m_{h}^{2}Na^{2}\beta }\sum_{\mathbf{k%
}}k^{2}\sum_{ip_{n}}G_{h}(\mathbf{k},ip_{n}+i\omega _{n})G_{h}(\mathbf{k}%
,ip_{n}) \\
\Pi ^{B}(i\omega _{n}) &=&-\frac{2e^{2}}{m_{h}^{2}N^{2}a^{4}\beta ^{2}}\sum_{%
\mathbf{k},\mathbf{q}}\sum_{ip_{n},ip_{n}^{\prime }}D^{A^{s}}(\mathbf{q}%
,ip_{n}^{\prime })G_{h}^{(0)}(\mathbf{k},ip_{n}-i\omega _{n})G_{h}^{(0)}(%
\mathbf{k}+\mathbf{q},ip_{n}+ip_{n}^{\prime })  \label{opc4}
\end{eqnarray}%
with $G(\mathbf{k},i\omega _{n})=1/\left[ i\omega _{n}-\xi _{\mathbf{k}%
}-\Sigma (\mathbf{k},i\omega _{n})\right] $.

After the frequency summation and analytic continuation, one finds%
\[
\mathrm{Im}\Pi ^{A}(\omega )=-\frac{e^{2}}{2m_{h}^{2}Na^{2}}\sum_{\mathbf{k}%
}k^{2}\int_{-\infty }^{\infty }\frac{d\varepsilon }{2\pi }%
A_{h}(k,\varepsilon )A_{h}(k,\varepsilon +\omega )[n(\varepsilon
)-n(\varepsilon +\omega )] 
\]%
where the spectral function $A_{h}(\mathbf{k},\omega )$ is defined as 
\[
A_{h}(\mathbf{k},\omega )=\frac{-2\Sigma ^{\prime \prime }(\mathbf{k},\omega
)}{\left[ \omega -\xi _{\mathbf{k}}-\Sigma ^{\prime }(\mathbf{k},\omega )%
\right] ^{2}+\left[ \Sigma ^{\prime \prime }(\mathbf{k},\omega )\right] ^{2}}
\]%
It is easy to find that without including $\Sigma (\mathbf{k},\omega )$, $%
\mathrm{Im}\Pi ^{A}(\omega )=0$ at finite $\omega $. The behavior of $\Sigma
^{\prime \prime }(\mathbf{k},\omega )$ has been discussed in the last
subsection. Its correction to $\mathrm{Im}\Pi ^{A}(\omega )$ turns out to be
in a higher-order as compared to the leading contribution in $\mathrm{Im}\Pi
^{B}(\omega )$ given below 
\begin{eqnarray}
\mathrm{Im}\Pi ^{B}(\omega ) &=&\frac{2e^{2}}{m_{h}^{2}N^{2}a^{4}}\sum_{%
\mathbf{k},\mathbf{q}}\int_{0}^{\infty }d\omega ^{\prime }\mathrm{Im}%
D^{A^{s}}(q,\omega ^{\prime })[n(\xi _{\mathbf{k}})-n(\xi _{\mathbf{k+q}%
})]\left\{ [n(\omega -\omega ^{\prime })+n(\omega ^{\prime })+1]\right. 
\nonumber \\
&&\left. \times \delta (\omega -(\xi _{\mathbf{k+q}}-\xi _{\mathbf{k}%
})-\omega ^{\prime })+[n(\omega ^{\prime })-n(\omega +\omega ^{\prime
})]\delta (\omega -(\xi _{\mathbf{k+q}}-\xi _{\mathbf{k}})+\omega ^{\prime
})\right\}  \label{opc9}
\end{eqnarray}%
\begin{figure}[tbp]
\begin{center}
\includegraphics[width=3.5in]
{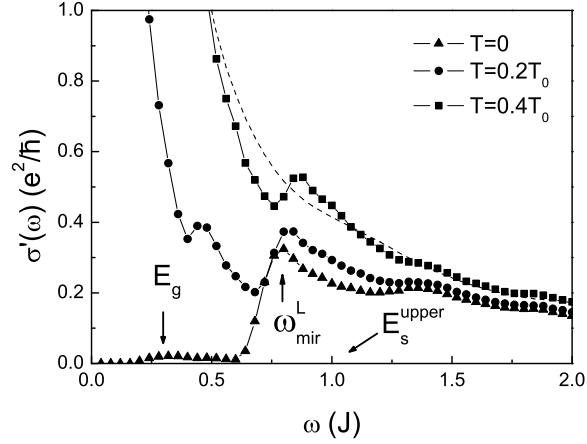}
\end{center}
\caption{Optical conductivity at different temperatures below $T_{0}$. A new
lower mid-infrared peak emerging at $\protect\omega _{\mathrm{mir}}^{L}$
which is weighted between the magnetic energy scales $E_{g}$ and $E_{s}^{%
\mathrm{upper}}$.}
\label{optical}
\end{figure}

To study the low-$\omega $ behavior, thus one needs only to keep the second
term: $\sigma ^{\prime }(\omega )\approx -\mathrm{Im}\Pi ^{B}(\omega
)/\omega $. In Fig. \ref{optical}, we have plotted the calculated optical
conductivity at various low temperature below $T_{0}$ at $\delta =0.125.$ A
prominent suppression of $\sigma ^{\prime }(\omega )$ at low-$\omega $ is
present at $T=0$ with a second \textquotedblleft
mid-infrared\textquotedblright\ peak emerging around $\omega _{\mathrm{mir}%
}^{L}\sim 0.75J$ which sits somewhat between the two characteristic magnetic
energy scales$,$ $E_{g}$ and $E_{s}^{\mathrm{upper}}$, as marked in the
figure. Note that such a new energy scale in the low-$\omega $ optical
conductivity merely reflects some weighted energy scale based on the
magnetic $\mathrm{Im}\chi ^{zz}$, which may be seen from the following
simplified formula for the optical conductivity at low temperature%
\[
\sigma ^{\prime }(\omega )\simeq \frac{2\pi ^{2}e^{2}N}{m_{h}^{2}N^{2}a^{4}%
\omega }\sum_{\mathbf{q}\neq 0}\int_{-\infty }^{\infty }d\omega ^{\prime }%
\frac{1}{a^{2}q^{2}}\mathrm{Im}\chi ^{zz}(q,\omega ^{\prime })(1+n(\omega
^{\prime }))\delta (\omega -\xi _{\mathbf{q}}-\omega ^{\prime }) 
\]%
With the increase of temperature, the \textquotedblleft
gap\textquotedblright\ at low energy in $\sigma ^{\prime }(\omega )$ is
quickly filled up by the thermal excitations as shown in Fig. \ref{optical}.
The lower \textquotedblleft mid-infrared\textquotedblright\ peak feature
remains around $\omega _{\mathrm{mir}}^{L}$ at low temperature throughout
the LPP below $T_{v}.$ Note that $T_{v}$ is between $T_{c}$ and $T_{0},$ and
the dashed curve at $T=0.4T_{0\text{ }}$is obtained by supposing that $%
T>T_{v}$ where $\mathrm{Im}\chi ^{zz}$ behaves differently.\cite{psnmr} As
compared to the solid curve at the same $T=0.4T_{0}$, which corresponds to
the case \emph{inside }the LPP, the overall difference is small except for
the vanishing the lower \textquotedblleft mid-infrared\textquotedblright\
peak [Fig. \ref{optical}].

Finally, we note that a Drude behavior without any signature of pseudogap
feature is indeed recovered if the phase at $T\geq T_{0}$ is \emph{%
extrapolated }to $T=0$ with using Eq. (\ref{static}):

\begin{equation}
\sigma ^{\prime }(\omega )\simeq -\frac{\mathrm{Im}\pi _{xx}^{B}(\omega )}{%
\omega }\rightarrow \frac{\delta e^{2}\left\langle \left( \Phi _{_{\square
}}^{s}\right) ^{2}\right\rangle }{2\omega ^{2}/t_{h}^{2}}
\end{equation}%
Of course, the perturbative approach is no longer expected to work reliably
at such strong flux fluctuations.

\section{summary and discussions}

\begin{figure}[tbp]
\begin{center}
\includegraphics[width=3.5in]
{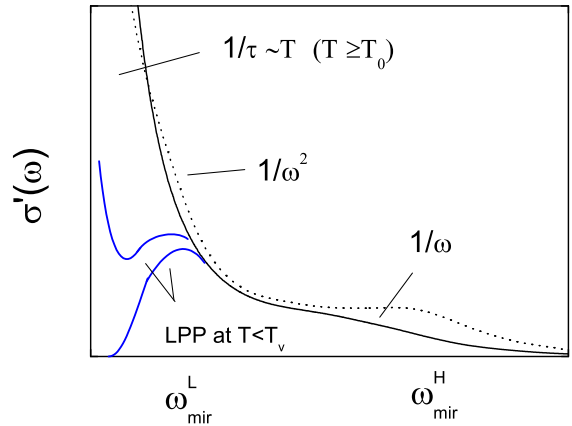}
\end{center}
\caption{A schematic optical conductivity summarizing the general behavior
at different energy and temperature regimes in the phase string model.}
\label{sum}
\end{figure}

We have studied the charge dynamics based on the phase string model of the
doped Mott insulator. This model possesses a mutual Chern-Simons gauge
interaction between the bosonic holons and spinons, which is responsible for
some highly non-trivial charge and spin dynamics in the system.

In the present work, the optical conductivity generally exhibits a
two-component feature: a \textquotedblleft coherent\textquotedblright\
low-energy part plus an \textquotedblleft incoherent\textquotedblright\
high-energy part. The former is Drude-like at high temperature, which
gradually gets suppressed, after entering the upper and lower pseudogap
phases, at $\omega <\omega _{\mathrm{mir}}^{L}.$ This is related to the
reduction of spin fluctuations, with a lower \textquotedblleft
mid-infrared\textquotedblright\ peak emerging at $\omega _{\mathrm{mir}}^{L}$
in $\sigma ^{\prime }(\omega )$ which is originated from a weighted magnetic
energy scale between $E_{g}$ (associated with the resonancelike energy in
the dynamic spin susceptibility) and $E_{s}^{\mathrm{upper}}$ (an upper
bound magnetic energy for $S=1$ excitations). The high-energy incoherent
part shows a resonance around $\omega _{\mathrm{mir}}^{H}\sim 4t_{h}$ if the
gauge fluctuation is sufficiently strong, which can smoothly evolve into a $%
1/\omega $ tail at weak gauge fluctuations, consistent with the situations
in underdoping and optimal/over-doping, respectively. The overall features
of the optical conductivity are summarized in Fig. \ref{sum}, whose behavior
is specified by two energy scales: $\omega _{\mathrm{mir}}^{L}$ and $\omega
_{\mathrm{mir}}^{H}$, which have been observed\cite{low1} experimentally. We
have also shown that the high mid-infrared resonance at $\omega _{\mathrm{mir%
}}^{H}$ can be closely correlated with the $\omega $ structure in the
density-density correlation function which agrees qualitatively with the
exact diagonalization results at all momenta.

The high-temperature Drude component is characterized by a linear-$T$
scattering rate $1/\tau _{\mathrm{dc}}$ in the region of $T>T_{0}$. This is
consistent with the previous result obtained\cite{pali3} using the quantum
Monte Carlo simulation for a phenomenological model of repulsive bosons
interacting with a strong spatially fluctuating gauge field. Furthermore, in
the same high-temperature regime, the scattering rate $1/\tau (\omega )$
also shows an approximate linear $\omega $ dependence in the case when the
high-$\omega $ optical conductivity evolves into a $1/\omega $ behavior.
Thus, the phase string model can give a consistent explanation for the
anomalous dc and ac transport in the normal state of the optimally doped
high-$T_{c}$ cuprates. Note that such linear $T$ or $\omega $ dependence of
the scattering rate above the pseudogap phase is really a high
temperature/energy behavior and thus is distinct from a quantum critical
phenomenon.

Physically, the phase boundary at $T_{0}$ separates the high-temperature
classical regime from the spin \textquotedblleft degenerate
regime\textquotedblright\ in the UPP where the spins form the RVB pairing
with $\Delta ^{s}\neq 0.$ In the classical regime, both the charge and spin
degrees of freedom behave diffusively, and the linear-$T$ relaxation rate
may be regarded as the system in the ``quantum limit of dissipation''.\cite%
{zaanen} It is interesting to see how small $T_0$ is as compared to a normal
Fermi degenerate temperature scale. The charge holons will gain quantum
coherence below a lower temperature $T_{v}$ in the \textquotedblleft boson
degenerate regime\textquotedblright\ as illustrated in Fig. \ref%
{phasediagram}, where the optical conductivity exhibits the pseudogap
behavior at low energy. Note that at $T\gtrsim T_{0}$, although $\Delta
^{s}=0,$ the superexchange $J$ still plays a critical role\cite{ps} to
\textquotedblleft repair\textquotedblright\ the spin \textquotedblleft $\hat{%
z}$-component mismatch\textquotedblright\ created by the hopping of the
holes to ensure an effective hopping integral, while the \textquotedblleft
irreparable\textquotedblright\ transverse spin mismatch described by the
phase string effect is responsible for the maximal gauge flux frustration
discussed in this paper. Such a \textquotedblleft normal
state\textquotedblright\ is expected to be smoothly connected to the
Brinkman-Rice retracable path limit\cite{br} at $T\gg T_{0}$, where the role
of $J$ eventually diminishes and the present effective model is no longer
valid.

Therefore, the simple boson model (\ref{md1}) seems to well capture many
important high-energy properties of the charge excitations in the cuprates
and the $t-J$ model. Another interesting channel to probe the high-energy
spin-charge separation is the ARPES experiment. We expect to see the
distinct energy scales in the single electron spectral function and a
detailed study of the high-energy structure based on the phase string model
will be presented elsewhere. It is further noted that in the present work we
have not considered the quasiparticle contribution to the optical
conductivity. In the phase string model, a spin-charge recombination will
occur in the superconducting phase in which nodal quasiparticles become
stable below the energy scale set by $E_{g}$.\cite{qp} But this is a fairly
low energy as compared to $\omega _{\mathrm{mir}}^{L},$ which should not
change the overall picture presented in Fig. \ref{sum}$.$

\bigskip \appendix

\section{spin dynamic susceptibility function}

According to Ref. \cite{psneutron}, the spinon Hamiltonian $H_{s}$ in Eq. (%
\ref{md3}) can be diagonalized as $H_{s}=\sum_{m\sigma }E_{m}\gamma
_{m\sigma }^{\dagger }\gamma _{m\sigma }+\mathrm{const.}$ by the Bogoliubov
transformation 
\begin{equation}
b_{i\sigma }=\sum_{m}w_{m\sigma }(i)[u_{m}\gamma _{m\sigma }-v_{m}\gamma
_{m-\sigma }^{\dagger }]  \label{bogoliubov}
\end{equation}%
with the spinon wave function $\omega _{m\alpha }\left( i\right) $
determined by the eigen equation: 
\begin{equation}
\xi _{m}\omega _{m\sigma }(i)=-J_{s}\sum_{j=nn(i)}e^{-i\sigma
A_{ji}^{h}}\omega _{m\sigma }(j)  \label{eig1}
\end{equation}%
and 
\begin{eqnarray}
u_{m} &=&\frac{1}{\sqrt{2}}\sqrt{\frac{\lambda }{E_{m}}+1}  \label{u} \\
v_{m} &=&\mathrm{sgn}(\xi _{m})\frac{1}{\sqrt{2}}\sqrt{\frac{\lambda }{E_{m}}%
-1}  \label{v}
\end{eqnarray}%
where the spinon excitation spectrum $E_{m}=\sqrt{\lambda ^{2}-\xi _{m}^{2}}$%
. Here the Lagrangian multiplier $\lambda $ is determined by enforcing the
average constraint $\left\langle \sum\nolimits_{\sigma }b_{i\sigma
}^{\dagger }b_{i\sigma }\right\rangle =1-\delta ,$ leading to 
\begin{equation}
2-\delta =\frac{1}{N}\sum_{m}\frac{\lambda }{E_{m}}\coth \frac{\beta E_{m}}{2%
}  \label{lam}
\end{equation}

Correspondingly the dynamic spin susceptibility function can be expressed as%
\cite{psneutron} 
\begin{eqnarray}
\chi ^{zz}\left( \mathbf{q},i\omega _{n}\right) &=&\frac{1}{N}%
\sum_{i,j}\int_{0}^{\beta }\left\langle S_{i}^{z}\left( \tau \right)
S_{j}^{z}\right\rangle e^{-i\mathbf{q}\cdot \left( \mathbf{r}_{i}-\mathbf{r}%
_{j}\right) +i\omega _{n}\tau }d\tau  \nonumber \\
&=&\frac{1}{4}\sum_{mm^{\prime }\alpha }C_{mm^{\prime }\alpha }\left( 
\mathbf{q}\right) \left[ \left( u_{m}u_{m^{\prime }}-v_{m}v_{m\prime
}\right) ^{2}\frac{n(E_{m^{\prime }})-n(E_{m})}{i\omega
_{n}+E_{m}-E_{m^{\prime }}}\right.  \nonumber \\
&&\left. +\left( u_{m}v_{m^{\prime }}-v_{m}u_{m\prime }\right) ^{2}\left(
1+n(E_{m})+n(E_{m^{\prime }})\right) \times \frac{1}{2}\left( \frac{1}{%
i\omega _{n}+E_{m}+E_{m^{\prime }}} -\frac{1}{i\omega
_{n}-E_{m}-E_{m^{\prime }}}\right) \right]  \label{sus}
\end{eqnarray}%
where $C_{mm^{\prime }\alpha }\left( \mathbf{q}\right) =\frac{1}{N}%
\sum_{ij}e^{-i\mathbf{q}\cdot (\mathbf{r}_{i}-\mathbf{r}_{j})}\omega
_{m\alpha }^{\ast }\left( j\right) \omega _{m\alpha }\left( i\right) \omega
_{m^{\prime }\alpha }^{\ast }\left( i\right) \omega _{m^{\prime }\alpha
}\left( j\right) $. By taking the analytic continuation $i\omega
_{n}\rightarrow \omega +i0^{+}$, one finally obtains 
\begin{eqnarray*}
\mathrm{Im}\chi ^{zz}\left( \mathbf{q},\omega \right) &=&-\frac{\pi }{2}%
\sum_{mm^{\prime }\alpha }C_{mm^{\prime }\alpha }\left( \mathbf{q}\right)
\{\left( u_{m}u_{m^{\prime }}-v_{m}v_{m\prime }\right) ^{2}\left[
n(E_{m^{\prime }})-n(E_{m})\right] \delta \left( \omega +E_{m}-E_{m^{\prime
}}\right) \\
&&+\left(u_{m}v_{m^{\prime }}-v_{m}u_{m\prime }\right) ^{2}\left[
1+n(E_{m})+n(E_{m^{\prime }})\right] \times \frac{1}{2}\left[ \delta \left(
\omega -E_{m}-E_{m^{\prime }}\right) -\delta \left( \omega
+E_{m}+E_{m^{\prime }}\right) \right] \}
\end{eqnarray*}

In the UPP at $T\geq T_{0}$, $J_{s}\rightarrow 0$ such that $E_{m}=\lambda $%
, $u_{m}=1$ and $v_{m}=0.$ $\lambda $ is determined by Eq. (\ref{lam}) as%
\cite{psnmr} $e^{\beta \lambda }=\left( 3-\delta \right) /\left( 1-\delta
\right) $. In this case, one finds the spin structure factor

\begin{eqnarray}
S^{zz}(\mathbf{q},\omega ) &=&\pi ^{-1}\left[ 1+n\left( \omega \right) %
\right] \mathrm{Im}\chi ^{zz}(\mathbf{q},\omega )  \nonumber \\
&=&\frac{1}{12}\left( 3-\delta \right) /\left( 1-\delta \right) \delta
(\omega )  \label{szz}
\end{eqnarray}%
which is concentrated at $\omega =0$ as there is no spin-spin correlation at
the mean-field level. Note that in obtaining the last line a correction
factor $2/3$ is multiplied such that the precise sum rule $\int d\omega
\sum_{\mathbf{q}}S^{zz}(\mathbf{q},\omega )=N\left( S^{z}\right) ^{2}=N/4$
is satisfied at half-filling.\cite{psneutron}

On the other hand, at $T=0$, due to the holon condensation, $A_{ji}^{h}$ can
be treated as describing a uniform flux satisfying 
\begin{equation}
\sum_{\mathrm{plaquette}}A_{ij}^{h}=\delta \pi
\end{equation}
which is expected to persist up to $T_{v}$. The detailed solution has been
presented in Ref. \cite{psneutron}. In Fig. \ref{chi}, the peak positions in 
$\mathrm{Im}\chi ^{zz}(\mathbf{q},\omega )$ at different doping
concentrations are shown.

\begin{acknowledgments}
We would like to thank W. Q. Chen, D. K. K. Lee, V. N. Muthukumar, X. L. Qi,
T. M. Rice, D. N. Sheng, J. Tu, and N. L. Wang for useful discussions. We
also acknowledge stimulating disucssions at Beijing Forum on
High-Temperature Superconductivity, which led to the present work. This work
has been partially supported by the NSFC grants.
\end{acknowledgments}

\end{document}